\documentclass[preprint,authoryear,12pt]{elsarticle}
\usepackage{graphicx}
\usepackage{amsmath}
\usepackage{mathtools}
\usepackage{amssymb}
\usepackage{natbib}
\usepackage[displaymath]{lineno}

 \biboptions{comma,sort}

\newcommand*\aap{A\&A}

\newcommand*\aj{AJ}

\newcommand*\apj{ApJ}

\newcommand*\grl{Geophys.~Res.~Lett.}

\newcommand*\icarus{Icarus}

\newcommand*\planss{Planet.~Space~Sci.}

\newcommand*\patchAmsMathEnvironmentForLineno[1]{%
  \expandafter\let\csname old#1\expandafter\endcsname\csname #1\endcsname
  \expandafter\let\csname oldend#1\expandafter\endcsname\csname end#1\endcsname
  \renewenvironment{#1}%
     {\linenomath\csname old#1\endcsname}%
     {\csname oldend#1\endcsname\endlinenomath}}% 
\newcommand*\patchBothAmsMathEnvironmentsForLineno[1]{%
  \patchAmsMathEnvironmentForLineno{#1}%
  \patchAmsMathEnvironmentForLineno{#1*}}%
\AtBeginDocument{%
\patchBothAmsMathEnvironmentsForLineno{equation}%
\patchBothAmsMathEnvironmentsForLineno{align}%
\patchBothAmsMathEnvironmentsForLineno{flalign}%
\patchBothAmsMathEnvironmentsForLineno{alignat}%
\patchBothAmsMathEnvironmentsForLineno{gather}%
\patchBothAmsMathEnvironmentsForLineno{multline}%
}

\journal{Planetary and Space Science}

\begin{document}

\begin{frontmatter}

\title{Librational response of a deformed 3-layer Titan perturbed by non-keplerian orbit and atmospheric couplings}

\author[imcce]{A. Richard\corref{cor1}}
\ead{arichard@imcce.fr}
\author[imcce,upmc]{N. Rambaux}
\author[upmc,lmd]{B. Charnay}

\address[imcce]{IMCCE, Observatoire de Paris, CNRS UMR 8028\\ 77 avenue Denfert-Rochereau, 75014 Paris, France}
\address[upmc]{Universit\'e Pierre et Marie Curie, UPMC - Paris 06}
\address[lmd]{Laboratoire de M\'et\'eorologie Dynamique, Paris}

\cortext[cor1]{Corresponding author: Phone:+33 [0]1 40 51 20 69}

\begin{abstract}

The analyses of Titan's gravity field obtained by Cassini space  mission suggest the presence of an internal ocean beneath its  icy surface.
The characterization of the geophysical parameters of the icy shell and the ocean is important to constrain the evolution models of  Titan. The knowledge of the librations, that are periodic oscillations  around a uniform rotational motion, can bring piece of information on  the interior parameters.

The objective of this paper is to study the librational response in  longitude from an analytical approach for Titan composed of a deep atmosphere, an elastic icy shell, an internal  ocean, and an elastic rocky core perturbed  by the gravitational interactions with Saturn.
We start from the librational equations developed for a rigid satellite in  synchronous spin-orbit resonance. We introduce explicitly the atmospheric torque acting on the surface computed from the Titan IPSL GCM \textit{(Institut Pierre Simon Laplace  General Circulation Model)} and the periodic deformations of elastic solid layers due to the tides. We investigate the librational response for various interior models in order to compare and to identify the  influence of the geophysical parameters and the impact of the elasticity.

The main librations arise at two well-separated forcing frequency ranges: low forcing frequencies dominated by the Saturnian annual and semi-annual frequencies, and a high forcing frequency regime dominated by Titan's orbital frequency around Saturn.
At low forcing frequency, the librational response is dominated by the Saturnian gravitational torque and the atmospheric torque has a small effect.
In addition, the libration amplitude in that case is almost equal to the magnitude of the perturbation. The modulation of the gravitational torque amplitude at the orbital frequency with periodic deformation induces long-period terms in the librational response which contain information on the internal structure.
At high forcing frequency the libration depends on the inertia of the layers and the elasticity can strongly reduce its amplitude at orbital frequency. For example, the amplitude of diurnal libration for oceanic models goes from about $320-390$ meters if the icy shell is purely rigid to $60-85$ meters when the elasticity is included, \textit{i.e.} a reduction of about $80\%$. For models without ocean, diurnal libration goes from $52$ meters in a rigid case to $50$ meters for an elastic case, a very low reduction due to the weak deformation of an entirely solid satellite compared to the deformation of a thin icy shell. Oceanic models with elastic solid layers have the same order of libration amplitude than the oceanless models, which makes more challenging to differentiate them by the interpretation of librational motion.
\end{abstract}


\end{frontmatter}

%\linenumbers

\section*{Introduction}

Titan exhibits a rich interior structure due to its large mean radius of 2575 kilometers. The recent measurements of the gravity field by \cite{Iess10, Iess12} reveal that Titan's moment of inertia (MoI) is as low as $0.33-0.34\ MR^2$. A large panel of internal structures, made of a low density core surrounded by icy layers, are consistent with this range of MoI and the mean density value of $1881$ kg m$^{-3}$ deduced by \cite{Iess10}.
In addition, the presence of a global internal ocean has been suggested from several techniques. 
First, \cite{Lorenz08} deduced the putative internal ocean through the response of the rotational motion of the surface to the atmospheric coupling.  
A second piece of evidence has been obtained from the determination of the non-zero obliquity by \cite{Stiles08} from the radar images of Cassini. They determined an obliquity equal to 0.3 degree that is larger than the $0.1$ degree obtained for a rigid Titan in a Cassini state (e.g. \cite{Bills08}). 
Then \cite{Bills08} and \cite{Baland11} suggested that this deviation is related to the influence of an internal ocean on the obliquity. 
Additional arguments in favor of this internal ocean have been obtained by electrical \citep{Beghin12}, topographical \citep{Nimmo10} and gravitational \citep{Iess12} analyses of Titan from Cassini-Huygens data.

Titan's rotational motion has been measured by \cite{Stiles08,Stiles10} that used the Cassini spacecraft's radar images in order to follow landmarks at the surface. 
They obtained a near-synchronous rotation for Titan with a drift rate of $0.00033$ degree per day, \textit{i.e.} $0.12$ degree per year \citep{Stiles10}. This approach has been recently revisited by \cite{Meriggiola12} that determined a synchronous spin-orbit motion within a residual of about 0.02 degree per year.
The main advantage of their approach is the introduction of Titan's librational motion in the reduction process. 
The librations describe the oscillations around the uniform rotational motion. Here we focus on the librations in longitude that correspond to the oscillations of the body principal axis projected onto the equatorial plane of the satellite. 
In the case of Titan, they have two distinct origins. The first one comes from the gravitational torque exerted by Saturn on the dynamical figure of Titan. 
The second one results from the coupling between the surface and the dense atmosphere of Titan, the atmospheric torque. Their amplitudes are modified by the interior parameters of Titan.
However, \cite{Goldreich10} suggested that the elastic behavior of the icy shell is not negligible and may strongly reduce the librational motion. Such prediction has been confirmed by \cite{VanHoolst13} that computed the libration in longitude at the orbital frequency
of Titan. In that case, Titan's surface will deform instead of rotate since the ocean figure should always point toward the planet and exert an elastic torque on the icy shell. 

The objective of this paper is to determine the librational reponse of an elastic Titan at various forcing frequencies resulting from its orbital motion. By including the atmospheric coupling,  we want to decorrelate surface forcing from the internal geophysical properties related to the internal ocean and perturbation of Titan's orbit. We investigate the wide spectrum of Titan's librations on contrary to \cite{Vanhoolst09,VanHoolst13} that focused on the diurnal frequency (Titan's orbital frequency) and its harmonics, and on the Saturnian semi-annual frequency (twice Saturn's orbital frequency) for the atmospheric coupling. The main interest is that in the orbital motion, there are librations at the Saturnian semi-annual frequency that comes from the interaction of Saturn with the Sun. Such librations have the same frequency than the main component of the atmospheric torque.  In addition, the orbital frequency spectrum can be in, or close to, resonance with some proper frequencies as shown for the Galilean satellites \citep{Rambaux11}.  Finally, the periodic variation of the gravitational torque amplitude at orbital frequency provides long-period terms in the libration with amplitudes dependent on the interior model.

In this paper, the recent interior models of Titan are also used (\textit{e.g.} \cite{Fortes12,Castillo10,McKinnon11}) in order to compute the values of the proper frequencies and to discuss the differences in the amplitude of librations. The elasticity is investigated by computing the radial deformations of surfaces due to the tides and responsible for the gravitational torques amplitude variations.

Finally, we use the 3D atmospheric model from \cite{LB12} that predicts an atmospheric torque smaller than in the \cite{Tokano05} paper used in all previous studies \citep{Lorenz08,Karatekin08,Vanhoolst09,VanHoolst13,Goldreich10}. 

In the first part of the paper, the internal structure models selected for the librational computation are described. The properties of these models depend on the history and energy sources of Titan (e.g. \cite{Tobie12}). We selected a broad range of possible internal scenarios for Titan in order to characterize the impact of the geophysical parameters on the librations.
The atmospheric torques of \cite{Charnay12} (called here CH12) and of \cite{Tokano05} (called here TO05) are described and discussed in Sect. 2.
The orbit of Titan is then analyzed in Sect. 3 by using the frequency analysis method providing the spectrum of the orbital motion. In Sect. 4, the elasticity is introduced in the librational equations through periodic torques amplitudes and the librations are analytically determined for the rigid and elastic cases.
Finally, the behavior of the libration angle at different forcing frequency ranges is analyzed and the influence of the geophysical parameters of the different interior models is discussed in Sect. 5.

\section{Interior models}
\label{sec:mis}

Titan's internal structure has been revealed by accurately tracking the trajectory of Cassini spacecraft approaching Titan during six flybys \citep{Iess10, Iess12}. They measured the gravity field and its variations allowing to infer information on the density profile of the satellite.
Since the inversion between the gravity field and density profile is not unique, only a range of models can be determined.
The add-on assumption that Titan is close to the hydrostatic equilibrium led \cite{Iess10} to obtain an estimation of the moment of inertia (MoI) $I$ of Titan between $0.33$ and $0.34\ MR^2$ (where $M$ and $R$ are the mass and mean radius of Titan, respectively). 
Such small MoI implies an increase of density towards Titan's center and requires a low-density core to match the mean density of $1881$ kg m$^{-3}$ \citep{Iess10}.

In parallel to these gravity measurements, two categories of thermal and chemical models have been developed to determine Titan's internal structure (\textit{e.g.} \cite{Castillo10,Castillo12, McKinnon11, Fortes12} and see the review of \cite{Tobie12}). In the first category, as developed by \cite{Fortes12}, the models described essentially the upper layers (dense and light ocean, comparison of solid models with pure water ice or methane clathrate layers),
while in the second category the models focused on the inner core composition assuming the presence of a global ocean layer (\textit{e.g.} \cite{Castillo12, McKinnon11}).
 
\cite{Castillo12} built interior models made of anhydrous silicate core surrounded by hydrated rock and water ice while  \cite{McKinnon11} focused on hydrated silicates surrounded by mixture of rock and ice. These models of inner core allow the presence of a small iron core, and are compatible with an internal ocean as deduced by \cite{Iess12}.

Six different models reported in Table \ref{table1} have been selected, coming from \cite{Fortes12} (the models called F1, F2 and the solid model F3), \cite{Castillo10} (models CA10 and FE10) and \cite{McKinnon11} (model MC11). The FE10 model is similar to the CA10 model with an additional small inner iron core as suggested by \cite{Castillo10}. The icy shell thickness has been taken equal to $100$ kilometers for each model. Such value corresponds to the upper bound obtained by topographical model of \citep{Nimmo10}. The lower bound has been obtained by \cite{Beghin12} by using the Schumann's resonance in Titan's atmosphere. The icy shell thickness is an essential parameter for Titan's models with a rigid shell \citep{Vanhoolst09} but the elasticity strongly diminish its influence (see Section \ref{sec:libration}). \cite{Fortes12} used oceans with a bottom mean radius of $2225$ kilometers (models F1 and F2). To be able to compare the influence of the ocean density and the inner core structure on the libration, the same ocean bottom mean radius is used for the CA10, MC11 and FE10 models. For a given set of solid layers sizes and densities, the ocean density is then adjusted to conserve Titan's MoI and mass. 

%\tar{The Love numbers $h_2$ quantify the deformation of the surface of those models under perturbing potentials. These quantities are obtained by numerical integration of the gravito-elastic equations of \cite{Alterman59}. The lowest value of $h_2$ is obtained for the entirely solid model F3 due to its expected lower deformation amplitude. The $h_2$ values for the different oceanic models is dependent on the interior structure parameters.}

%===
% TABLE INTERIOR MODELS
%==
\begin{table*}[]
\footnotesize
\caption{\label{table1} Characteristics of the internal structure models selected for this paper. The first three models quoted \textit{Light Ocean} (F1), \textit{Dense Ocean} (F2) and \textit{Pure-water-ice} (F3) models are from \cite{Fortes12}. Model CA10 is from \cite{Castillo10} and MC11 from \cite{McKinnon11}, both with dense ocean and high pressure ice layers. The last model (FE10) is a model similar to CA10 including an iron core. $I$ is the total moment of inertia of the body, while $C_s$ and $C_i$ are polar moments of inertia of the icy shell and solid interior respectively, expressed in terms of mass and radius of Titan ($M_T=13455.3\ 10^{19}$ kg and $R_T=2575.5$ km following \cite{Fortes12}).}
\begin{center}
\begin{tabular}{l c c c c c c}
 \hline
  & \multicolumn{2}{c}{\begin{bf} F1 \end{bf}}  & \multicolumn{2}{c}{\begin{bf} F2 \end{bf} } &  \multicolumn{2}{c}{\begin{bf} F3 \end{bf}}\cr
  \hline
  Layer & $\rho$ (kg m$^{-3}$)  & R (km) & $\rho$ (kg m$^{-3}$) & R (km)& $\rho$ (kg m$^{-3}$) & R (km)\cr
  \hline
  Icy shell & 930.9 & 2575.5 & 930.9 & 2575.5 &  932.8 & 2575.5 \cr
  Ocean & 1023.5 & 2475 & 1281.3 & 2475 & - & - \cr
  HP ice 1 & - & - & - & - & 1193.3 & 2429  \cr
  HP ice mantle & 1272.7 & 2225 & 1350.9 & 2225 & 1268.7 & 2304  \cr
  Silicate mantle & 1338.9 & 2163 & - & - &1338.9 & 2185 \cr
  Silicate core & 2542.3 & 2116 & 2650.4 & 1984 & 2584.1 & 2055\cr
\hline
  $I$ (MR$^2$) & \multicolumn{2}{c}{0.3414}&\multicolumn{2}{c}{0.3413}&\multicolumn{2}{c}{0.3413}\cr
  $C_s$ (MR$^2$) & \multicolumn{2}{c}{0.0357}&\multicolumn{2}{c}{0.0357}&\multicolumn{2}{c}{-}\cr
  $C_i$ (MR$^2$) & \multicolumn{2}{c}{0.2320}&\multicolumn{2}{c}{0.2133}&\multicolumn{2}{c}{-}\cr
  \hline
  \cr

  & \multicolumn{2}{c}{\begin{bf} CA10 \end{bf}}  & \multicolumn{2}{c}{\begin{bf} MC11 \end{bf} } &  \multicolumn{2}{c}{\begin{bf} FE10 \end{bf}}\cr
  \hline
  Layer & $\rho$ (kg m$^{-3}$)  & R (km) & $\rho$ (kg m$^{-3}$) & R (km)& $\rho$ (kg m$^{-3}$) & R (km)\cr
  \hline
  Icy shell & 932.8 & 2575.5 & 932.8 & 2575.5 &  932.8 & 2575.5 \cr
  Ocean & 1317.6 & 2475 & 1315.5 & 2475 & 1224.1 & 2475 \cr
  HP ice mantle & 1350.9 & 2225 & 1350.9 & 2225 & 1350.9 & 2225  \cr
  Rocky mantle & 2520 & 2000 & 2200 & 2075 & 2520 & 2000 \cr
  Silicate core & 3400 & 900 & 3300 & 1300 & 3400 & 900\cr
  Iron core & - & - & - & - & 6500 & 500 \cr
        \hline
  $I$ (MR$^2$) & \multicolumn{2}{c}{0.3402}&\multicolumn{2}{c}{0.3378}&\multicolumn{2}{c}{0.3336}\cr
  $C_s$ (MR$^2$) & \multicolumn{2}{c}{0.0358}&\multicolumn{2}{c}{0.0358}&\multicolumn{2}{c}{0.0358}\cr
  $C_i$ (MR$^2$)& \multicolumn{2}{c}{0.2095}&\multicolumn{2}{c}{0.2073}&\multicolumn{2}{c}{0.2096}\cr
  \hline
\end{tabular}
\end{center}
\end{table*}

\section{Atmospheric torque}
\label{sec:atmo}

Titan has a thick atmosphere extending up to 800 km with a surface pressure of nearly 1.5 bars. Exchanges of angular momentum happen between the atmosphere and the surface, producing an atmospheric torque influencing Titan's rotation. The atmosphere dynamic is mainly driven by insolation, which creates a circulation zone called a Hadley cell between hotter and cooler regions. The circulation in Titan's troposphere is essentially dominated by one Hadley cell extending from one pole to the other. At the equinox, the circulation reverses with the formation of two Hadley cells, and the ascending air zone (also called the Intertropical Convergence Zone) moves from one pole to the other. This reversal produces a seasonal angular momentum exchange with Saturn's semi-annual period ($5376.633$ terrestrial days) and has been suggested as the main torque influencing Titan's rotation \citep{Tokano05}.
The total atmospheric angular momentum (AAM) is given by \citep{Tokano05}

\begin{equation}
l_{atm}=\frac{2\pi R^3}{g}\int^{p_s}_{0} \int^{\pi/2}_{-\pi/2}u\cos^2\psi d \psi d p,
\end{equation}
where $R=2575.5$ km is Titan's mean radius, $g= 1.354$ m$^{-2}$s$^{-1}$ is Titan's surface gravity, $p_{s}$ is the surface pressure, $u$ is the zonal wind speed (i.e. the horizontal wind speed in the west-east direction and relative to the surface, $u$ is positive for eastward wind), $\psi$ is the latitude and $p$ is the pressure. 

The atmospheric torque is then
\begin{equation}
\Gamma_{a}=- \frac{dl_{atm}}{dt}.
\end{equation}

The zonal wind speed of an air parcel in the ascending zone is approximately null due to a strong surface friction. Its angular momentum is conserved during its meridional transport from the ascending zone to the poles. Thus, the AAM is maximum at equinoxes (i.e. when the ascending zone is the farthest from Titan's rotation axis), and minimum at solstices (i.e. when the ascending zone is the nearest to Titan's rotation axis) \citep{Tokano05}. 
This change corresponds to an exchange of angular momentum with the surface by friction of winds.

The angular momentum exchange, and so the atmospheric torque, depends essentially on the extension of the Hadley cell in latitude and altitude \citep{Mitchell09}.
All previous studies on Titan's non-synchronous rotation \citep{Lorenz08,Karatekin08,Vanhoolst09, VanHoolst13,Goldreich10} used the atmospheric torque at Saturnian semi-annual frequency from the general circulation model (GCM) of \cite{Tokano05} ($\Gamma_A=1.6\ 10^{17}$ kg m$^2$ s$^{-2}$; this torque is called here TO05). In this study, the atmospheric torque from the Titan IPSL GCM (Institut Pierre-Simon Laplace General Circulation Model) \citep{LB12} is also used. This model successfully reproduces the winds and the thermal structure observed by Cassini and Huygens \citep{LB12,Charnay12}. \cite{Charnay12} show that the Hadley cell is essentially trapped in the first two kilometers. This trapping, in addition to a lower latitudinal extension  of the cell, reduces the angular momentum exchange by approximately a factor 10 compared to \cite{Tokano05}. 

The latitudinal extension of the Hadley cell depends a great deal on the thermal inertia of the ground (\textit{i.e.} the ground thermal response to solar heating).
Its value for Titan is unknown, it has been estimated to be around $340$ J m$^2$ s$^{-1/2}$ K$^{-1}$ \citep{Tokano05b} and used in \cite{Tokano05}, but the value may be higher. Atmospheric torques have been calculated with a thermal inertia of $400$ and $2000$ J m$^2$ s$^{-1/2}$ K$^{-1}$, the last one corresponding to a realistic maximal value. In that case, the Saturnian semi-annual frequency component of the Titan IPSL GCM torque (called here CH12) is of about $\Gamma_A=2\ 10^{16}$ kg m$^2$ s$^{-2}$ and is reduced to $\Gamma_A=1.1\ 10^{16}$ kg m$^2$ s$^{-2}$ for a model with a thermal inertia of $2000$ J m$^2$ s$^{-1/2}$ K$^{-1}$. In this study, we focus on the first model providing the largest Saturnian semi-annual frequency torque amplitude.

\begin{figure}[!htbp]
 \centering
 \resizebox{\hsize}{!}{
\includegraphics[width=0.6\textwidth,clip]{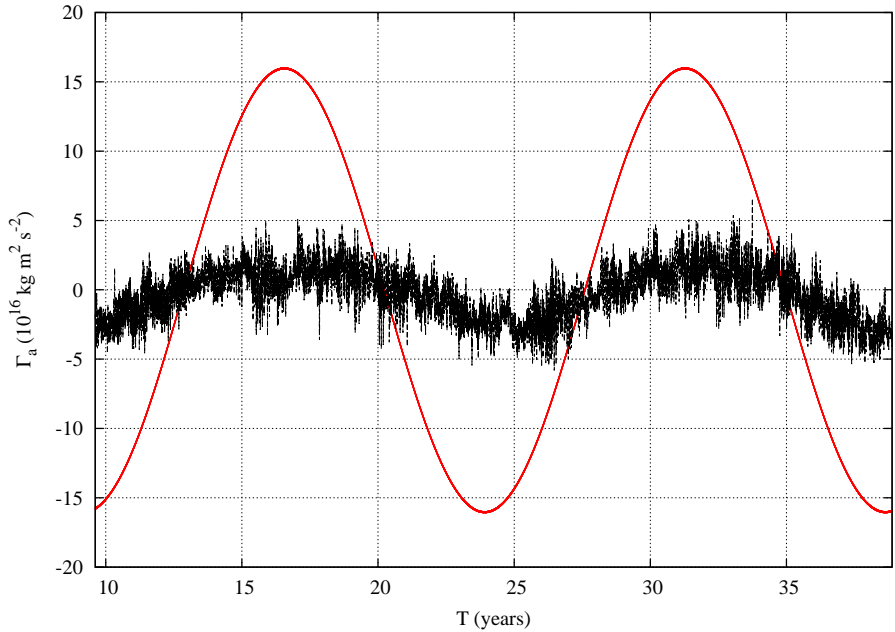}}     
%%% Note the ABSENCE of the extension .pdf , .eps or .ps  !
  \caption{Atmospheric torque amplitude along the Z-axis for the Titan IPSL GCM torque (dark plot) with a thermal inertia of $400$ J m$^2$ s$^{-\frac{1}{2}}$ K$^{-1}$. The amplitude of the Saturnian semi-annual frequency torque TO05 is plotted in red for comparison. Initial date is J2000.}
  \label{atmocompare}
\end{figure}

 The atmospheric torque CH12 with a thermal inertia of $400$ J m$^2$ s$^{-1/2}$ K$^{-1}$ and the Saturnian semi-annual frequency torque TO05 are represented in Fig. \ref{atmocompare}. Many other frequencies than the Saturnian semi-annual one appear in the spectrum of the atmospheric torque CH12. The most important ones are frequencies corresponding to $1$ (linked to the eccentricity of the orbit of Saturn around the Sun), $1/3$  and $1/4$ (other harmonics from the seasonal cycle) Saturnian year with amplitudes below $5\ 10^{15}$ kg m$^2$ s$^{-2}$. The other frequencies correspond to atmospheric waves that are either free (baroclinic waves, kelvin waves,...) or forced by the thermal tides and the gravitational tides caused by Saturn \citep{Tokano02}. The free atmospheric waves are very dependent on the model, but their amplitudes appear to be small. Concerning the thermal and gravitational tides, they have a very limited impact on the tropospheric winds and therefore on the exchange of angular momentum with the surface in the model from \cite{LB12}. Other effects may influence the atmospheric torque. \cite{Mitchell09} suggested the methane cycle could reduce the amplitude of the torque and produce a small drift while \cite{Tokano12} suggested the presence of mountains could increase the AAM a little without changing the torque very much. These effects remain poorly constrained but should only have a small impact.

\section{Orbital forcing}
\label{sec:orbit}

The librational motion is the rotational response of Titan to the gravitational or atmospheric torques exerted on its dynamical figure. 
The gravitational torque depends on the relative distance between Titan and Saturn and on the true longitude of Titan (\textit{e.g.} \cite{Rambaux11}). The orbital elements are time-dependent and can be approximated by quasi-periodic Fourier series resulting from the interaction of Titan with Saturn, the Sun or Iapetus \citep{Vienne95}.
 
Here, we develop a linear theory of the librational motion of Titan that is approximated in small quantities, the amplitude of libration and the eccentricity. The orbital motion of Titan comes from the JPL $Horizons$ ephemerides that contain the last observation from Cassini space mission \citep{Giorgini96}.
The ephemerides are given over the period of 07-Jan-1800 to 07-Jan-2200 corresponding to an interval of 400 years. The frequency analysis method developed by  \cite{Laskar88,Laskar03} and implemented in the TRIP software \citep{Gastineau12} is used in order to decompose the true longitude into Fourier series.

\begin{figure}[!htbp]
 \centering
\includegraphics[scale=0.45]{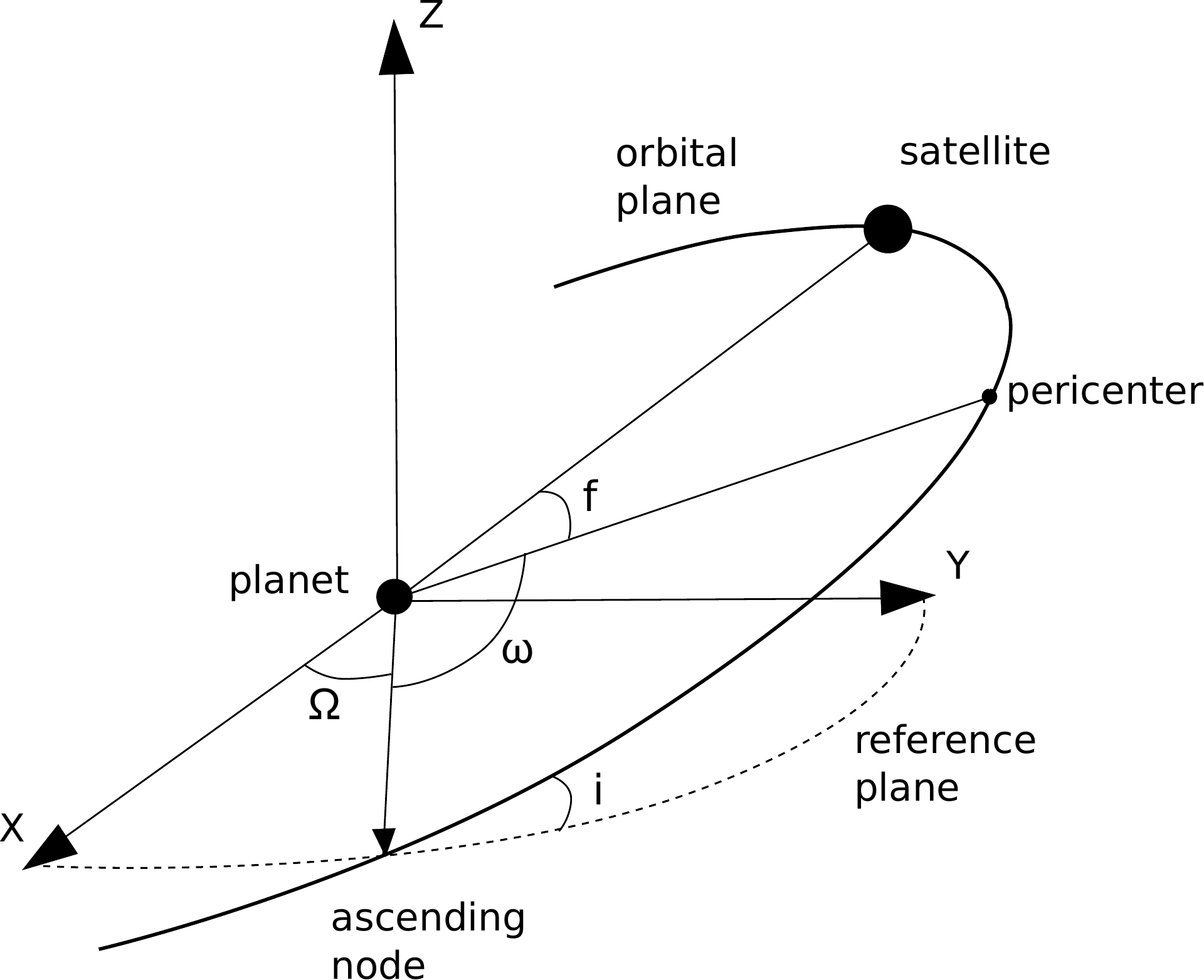}     
%%% Note the ABSENCE of the extension .pdf , .eps or .ps  !
  \caption{The geometry of the satellite orbit around the planet with respect to the reference plane.}
  \label{figureorbit}
\end{figure}

The decomposition of the true longitude $\nu$ is given in Table \ref{table2}. The notation of \cite{Vienne95} is used, where $L_6$ represents the mean longitude of Titan ($L=\Omega+\omega+M$ with $\Omega$ the longitude of node, $M$ the mean anomaly and $\omega$ the argument of pericenter, see Figure \ref{figureorbit}), $\varpi_6$ and $\varpi_8$ the longitudes of pericenter ($\varpi=\Omega+\omega$) of Titan and Iapetus, respectively, $\Omega_6$ and $\Omega_8$ the longitudes of node of Titan and Iapetus, and $L_s$ the mean longitude of the Sun. The true longitude $\nu$ is given by $\nu=\Omega+\omega+f$ with $f$ the true anomaly.

The largest term of the Table \ref{table2} corresponds to the first term of the development of the difference between the true anomaly $f$ and the mean anomaly $M$ as function of the eccentricity $e$, \textit{i.e.} $2e\sin M$. The magnitude is equal to 11899.3237" corresponding to an eccentricity of 0.0288 and the frequency is equal to the orbital frequency. The following term corresponds to the second term of the development in the true anomaly, \textit{i.e.} $\frac{5}{4}e^2\sin 2M$ and oscillates at twice the orbital frequency. 
The other frequencies result from the motion of Saturn around the Sun ($L_s$, $2L_s$, and $3L_s$ here called Saturnian annual, semi-annual and ter-annual terms respectively, and the combination $L_6+\Omega_6-2L_s$) and interaction with Iapetus ($L_6-2\varpi_8+2\Omega_6$) (see \cite{Vienne95}). The last term has a frequency close to the combination of ($12L_{j}-13L_{s}$) where $L_j$ is the mean longitude of Jupiter.
The last two terms of the Table~\ref{table2} with periods of $3583$ and $640$ days respectively are small terms of the true longitude series with magnitudes below $10 "$. These terms have small magnitudes but due to their frequency values, the librational response amplitudes are greater than the 10 meters limit which is used for the Table \ref{table4} (see Section \ref{sec:libration}). 

Let us note that the frequency analysis of the orbit over $400$ years is not sufficient to identify the periods longer than several hundred years. \cite{Vienne95} have identified longer periods in Titan's orbital motion listed in the $TASS$ analytical ephemerides. 
Frequencies associated with $\Omega_8$ ($3263.07$ years), $\varpi_8$ ($3181.86$ years), $\Omega_6$ ($703.51$ years) and $\varpi_6$ ($703.30$ years) possess high magnitudes. Since they are the signature of time variations much larger than the 400 years of the $Horizons$ ephemerides, these terms act like secular deviation on the true longitude. 
These secular terms are removed from the analysis in the determination of the linear deviation of the true longitude. 
However, for a longer time range, the long period terms from longitudes of nodes and pericenters have to be considered explicitly.

%===
% TABLE FREQUENCY ANALYSIS
%==
\begin{table*}[]
\footnotesize
\caption{\label{table2} Frequency analysis of the true longitude of Titan from $Horizons$ ephemerides. The difference between true longitude $\nu$ and mean anomaly $M$ is decomposed according to $\nu-M-\phi_0=\sum_jH_j \sin(\omega_jt+\alpha_j)$, with $\phi_0$ is the initial value of the satellite rotation angle measured from the line of ascending node, $H_j$ the magnitude, $\omega_j$ the frequency and $\alpha_j$ the phase. The frequency of the last term is close to the combination of ($12L_{j}-13L_{s}$) where $L_j$ is the mean longitude of Jupiter. This table presents the main terms of the true longitude series that acts on the librational motion. Initial date is J2000.}
\begin{center}
\begin{tabular}{l r c c r}
 \hline
\begin{bf}Freq. \end{bf}&\begin{bf} Period \end{bf}  &\begin{bf} Magnitude \end{bf}  &  \begin{bf} Phase \end{bf} &  \begin{bf} Identification \end{bf}\\
\begin{bf}(rad/days) \end{bf}&\begin{bf} (days)\end{bf}  &\begin{bf} ('') \end{bf}  &  \begin{bf} (degree) \end{bf} &  \\
  \hline
0.394018 &     15.9464 &  11899.3237 &    163.3693 & $L_6-\varpi_{6}$\cr 
0.788036 &      7.9732 &    212.5868 &    -32.7941 &$2L_6-2\varpi_{6}$ \cr  
0.394081 &     15.9439 &     56.6941 &    -68.1211 &$L_6-2\varpi_{8}+2\Omega_{6}$\cr 
0.001169 &   5376.6331 &     43.7313 &    -66.0428 &$2L_{s}$\cr  
0.000584 &  10750.3648 &     37.5508 &    138.4821 &$L_{s}$\cr 
0.392897 &     15.9919 &     31.5673 &     10.8789 &$L_6+\Omega_{6}-2L_{s}$\cr
0.001753 &   3583.9304 &     5.6147  &    250.1412 & $3L_{s}$ \cr
0.009810 &    640.4892 &      1.4983 &   -77.2905 & - \cr
        \hline
\end{tabular}
\end{center}
\end{table*}

\section{Librational model}

\subsection{Rigid case}
\label{sec:rigid}

The librational equations describing the variations of Titan's rotation are obtained from the angular momentum equation applied to each layer of the satellite. In the case of rigid solid layers, the hydrostatic equilibrium shape is composed of a static tidal bulge on which the planet exerts a gravitational forcing at each instant. In addition, the misalignment of the static tidal bulges of the shell and the inner core generates an internal gravitational torque coupling the upper layers to the inner core.
Following the developments of \cite{Vanhoolst08, Vanhoolst09} and \cite{Rambaux11}, the equations governing the librational motions of the rigid triaxial shell and inner core perturbed by external and internal gravitational forces are written as
\begin{equation}
\begin{cases}
C_s\ddot{\gamma_s}+(K_s+2K_{int})\gamma_s -2K_{int}\gamma_i =K_s(\nu-M-\phi_{o,s}),\\
C_i\ddot{\gamma_i}+(K_i+2K_{int})\gamma_i -2K_{int}\gamma_s =K_i(\nu-M-\phi_{o,i}),
 \label{eqlib}
 \end{cases}
\end{equation}
where subscripts $s$ and $i$ refer to the icy shell and to the inner core respectively, $\gamma$ is the libration angle, $M$ the mean anomaly,  $\nu$ is the true longitude, $\phi$ is the rotation angle of the satellite's longest axis measured from the line of the ascending node and $\phi_{o,l}$ its initial value for the layer $l$ (where subscript $l$ refers to the shell or the inner core), $n$ is the mean motion. $K_s$ and $K_i$ are the amplitudes of the effective gravitational torques exerted by Saturn on the dynamical figure of each layer, while $K_{int}$ is the amplitude of the internal gravitational torque exerted by the shell on the inner core due to their misalignment. The gravitational torques amplitudes are given by $K_s=3n^2[(B_s-A_s)+(B_s'-A_s')]$ and $K_i=3n^2[(B_i-A_i)-(B_i'-A_i')]$, where $A_l$, $B_l$ and $C_l$  are the principal moments of inertia (defined as $C_l>B_l>A_l$) and $A_l'$, $B_l'$ and $C_l'$ are the ocean pressure effect on the solid layer $l$ expressed as increment of inertia. The internal gravitational torque amplitude $K_{int}$ also depends on the geophysical parameters of the body as described by \cite{Vanhoolst09}.
In this study, the dynamical equations are linearized as function of the eccentricity and libration amplitude. 
The frequencies of the forcing torques acting on the two solid layers are given by the Fourier series of $\nu-M$ developed in the previous section.

The proper frequencies (or natural frequencies) of the satellite are described in terms of free frequencies of the harmonic oscillator as done by \cite{Dumberry11}. First, the free frequencies corresponding to the system (\ref{eqlib}) free of internal gravitational torque ($K_{int}=0$) are 
\begin{eqnarray}
\omega_s =\sqrt{\frac{K_s}{C_s}}= n\sqrt{\frac{3[(B_s-A_s)+(B_s'-A_s')]}{C_s}}\ , \label{omegas2}\\
\omega_i =\sqrt{\frac{K_i}{C_i}}= n\sqrt{\frac{3[(B_i-A_i)-(B_i'-A_i')]}{C_i}}\ .
\end{eqnarray}
Then, the free mode of oscillation corresponding to the system (\ref{eqlib}) without Saturn's gravitational torque ($K_s=K_i=0$) is expressed as
\begin{equation}
\mu=\sqrt{2K_{int}\Bigl(\frac{1}{C_i}+\frac{1}{C_s}\Bigl)}\ .
\end{equation}
Finally, the proper frequencies of the system (\ref{eqlib}) forced by Saturn and internal gravitational torques can be written as
\begin{eqnarray}
\omega_1^2 =\frac{1}{2}(\omega_i^2+\omega_s^2+\mu^2 + \sqrt{\Delta})\ , \label{omega1}\\
\omega_2^2 =\frac{1}{2}(\omega_i^2+\omega_s^2+\mu^2 - \sqrt{\Delta})\ , \label{omega2}
\end{eqnarray}
with $\Delta=\omega_s^4+\omega_i^4+\mu^4+2\omega_i^2\mu^2+2\omega_s^2\mu^2-2\omega_i^2\omega_s^2-8\omega_i^2\frac{K_{int}}{C_s}-8\omega_s^2\frac{K_{int}}{C_i}$. The values of $\omega_1$ and $\omega_2$ are given for each selected model in Table \ref{table3}.

Using these expressions of the proper frequencies, the solutions of the system (\ref{eqlib}) are written as $\gamma_l=\sum_j \gamma_{l,j} \sin(\omega_j t+\alpha_j)$  by setting $\nu-M-\phi_{o,l}=\sum_j H_j\sin(\omega_j t+\alpha_j)$, where $\gamma_{l,j}$ and $H_j$ are respectively the amplitude of libration and the magnitude of perturbation, $\omega_j$ is the frequency of perturbation and $\alpha_j$ the phase. The icy shell libration amplitudes for an oceanic model are then given by
\begin{eqnarray}
\gamma_{s,j}=\frac{1}{C_iC_s}\frac{H_j[K_s(K_i+2K_{int}-\omega_j^2 C_iK_s)+2K_{int}K_i]}{(\omega_j^2 - \omega_1^2)(\omega_j^2 - \omega_2^2)}\ . %=F(\omega_j) H_j\  .
 \label{gamma}
\end{eqnarray}
The libration amplitudes for a rigid solid model are given in Appendix.

As seen in Sect. \ref{sec:atmo}, the atmospheric torque exerted on Titan's surface is acting like a forcing term with a main component at Saturnian semi-annual frequency and many components with amplitudes below $5\ 10^{15}$ kg m$^2$ s$^{-2}$.
The atmospheric torque is introduced into the right-hand side of the first equation of the system (\ref{eqlib}) as $H_jK_s\sin(\omega_jt+\alpha_j)+\Gamma_{A,j}\sin(\omega_jt+\alpha_j+\Delta\alpha_j)$, where $\Gamma_{A,j}$ is the atmospheric torque magnitude, $\omega_j$ the frequency and $\Delta\alpha_j$ the phase difference with $\alpha_j$. The libration amplitude is then decomposed into a sine and a cosine term expressed as
\begin{eqnarray}
 \label{ampsin}
\gamma^s_{s,j}=\frac{1}{C_iC_s}\frac{H_j[K_s(K_i+2K_{int}-\omega_j^2 C_iK_s)+2K_{int}K_i]}{(\omega_j^2-\omega_1^2)(\omega_j^2-\omega_2^2)}\\+\frac{\Gamma_{A,j}\cos(\Delta\alpha_j)(2K_{int}+K_i-\omega_j^2C_i)}{C_iC_s(\omega_j^2-\omega_1^2)(\omega_j^2-\omega_2^2)}\ \nonumber,
\end{eqnarray}
\begin{equation}
\gamma^c_{s,j}=\frac{1}{C_iC_s}\frac{\Gamma_{A,j}\sin(\Delta\alpha_j)(2K_{int}+K_i-\omega_j^2 C_i)}{(\omega_j^2-\omega_1^2)(\omega_j^2-\omega_2^2)}\ ,
\label{ampcos}
\end{equation}
where $\gamma^s_{s,j}$ and $\gamma^c_{s,j}$ are the amplitudes of the sine and cosine terms respectively of the libration under orbital and atmospheric influence, as previously $\omega_j$ is the perturbing frequency, $\alpha_j$ its phase, and $\Delta\alpha_j$ the phase difference with the orbital perturbation. The cosine term implies a difference of phase with the libration amplitude due to orbital perturbation. For frequencies of the atmospheric torque which are not present in the orbital motion, the librational motion is obtained by doing $H_j=0$ in Eq.~(\ref{ampsin}).

\subsection{Elastic case}

The elasticity is introduced by modeling the radial deformations of the surfaces of the satellite layers and the resulting variations of the moments of inertia are computed. 
Each surface composing the satellite is parametrized by $r(r_0,\theta,\lambda)=r_0 + u_r(r_0,\theta,\lambda)$ where $r_0$ is the mean radius of the equivalent sphere and the angles $\theta$, $\lambda$ are the colatitude and longitude, respectively. Then $u_r$ is the radial deformation induced by the centrifugal and tidal potential.

\cite{Hinderer82} used a decomposition of $u_r$ into spherical harmonics. At the second order, we have
\begin{equation}
u_r(r_0,\theta,\lambda)=r_0(\sum_{j=0}^2 d_{2,j}P_{2}^j(\cos\theta)\cos j\lambda + e_{2,j}P_{2}^j(\cos\theta)\sin j\lambda),
\label{defdij}
\end{equation}
where $d_{2,j}$ and $e_{2,j}$ quantify the radial deformation normalized to $r_0$ and $P_{i}^j(\cos \theta)$ are associated Legendre polynomials. By using the definition of $r$, the non-zero components of the inertia tensor $[I]$ can be written at first order in terms of $d_{2,j}$, $e_{2,j}$
\begin{align}
I_{11}&=\frac{8\pi}{3}\int_0^R\rho(r)\Bigl(r^4+\frac{1}{10}\frac{d}{dr}(d_{20}(r)r^5)-\frac{3}{5}\frac{d}{dr}(d_{22}(r)r^5)\Bigl)dr,\\
I_{22}&=\frac{8\pi}{3}\int_0^R\rho(r)\Bigl(r^4+\frac{1}{10}\frac{d}{dr}(d_{20}(r)r^5)+\frac{3}{5}\frac{d}{dr}(d_{22}(r)r^5)\Bigl)dr,\\
I_{33}&=\frac{8\pi}{3}\int_0^R\rho(r)\Bigl(r^4-\frac{1}{5}\frac{d}{dr}(d_{20}(r)r^5)\Bigl)dr,\\
I_{12}&=-\frac{8\pi}{15}\int_0^R\rho(r)\frac{d}{dr}(e_{22}(r)r^5)dr.
\end{align}

\cite{Love11} defined at the surface the Love number $h_2$ which characterizes the body response to second order perturbing potential. Here, we use the following definition for the radial deformation $u_r$
\begin{equation}
u_r(r_0,\theta,\lambda)=\frac{H(r_0)}{g(R)}V_2 \ ,
\label{raddef}
\end{equation}
 where $g(R)$ represents the surface gravity of Titan and $V_2$ is the perturbing potential developed at order $2$. $H$ is the dimensionless radial function corresponding to $h_2$ at the surface. By using eqs. (\ref{defdij}) and (\ref{raddef}), a straight relation exists between the $d_{2,j}$, $e_{2,j}$ and the response to the perturbing potential proportional to $H$.

As shown by \cite{Giampieri04}, the development at first order in eccentricity of the perturbing potential (centrifugal and tidal) can be decomposed into a secular and a periodic part
\begin{multline}
V_2=\frac{Gm}{R}\Bigl(\frac{r}{R}\Bigl)^2q_t\Bigl[\frac{1}{6}P_{2}^0(\cos \theta)-\frac{1}{12}P_{2}^2(\cos \theta)\cos 2\lambda \\+ e \Bigl[\frac{1}{2}P_{2}^0(\cos \theta)\cos M - \frac{1}{3}P_{2}^2(\cos \theta)\sin 2\lambda \sin M\\ -\frac{1}{4}P_{2}^2(\cos \theta)\cos 2\lambda \cos M)\Bigl]\Bigl],
\end{multline} 
where $m$ and $R$ are the mass and mean radius of the satellite, and $q_t=-3\frac{M_p}{m}\Bigl(\frac{R}{a}\Bigl)^3$ with $M_p$ the mass of the planet and $a$ the semi-major axis of the satellite. The centrifugal potential has been supposed constant with a periodic variation of the potential $V_2$ only due to tides.

The secular part of the potential corresponds to static tides and is responsible for the static bulge of the satellite which is already included in our rigid case (see Sect. \ref{sec:rigid}). The periodic part, which is of order of the eccentricity $e$, governs time-dependent deformation of the surfaces.

The deformations of the satellite layers due to the tides induce a time-dependent gravitational potential. Mass distribution varies periodically and the inertia tensor of the body can be decomposed into a static part $I^s$ (corresponding to the static potential) and a periodic part $I^p$ associated to deformations around the static bulge.
Using the definitions of $u_r$ and of the potential $V_2$, the periodic components of the inertia tensor are given by
 \begin{eqnarray}
I_{11}^p= - \Delta I e \cos M,\\
I_{22}^p=\frac{\Delta I}{2} e\cos M , \\
I_{33}^p=\frac{\Delta I}{2} e\cos M,\\
I_{12}^p=-\Delta I e \sin M,\\
I_{13}^p=I_{23}^p=0,
\end{eqnarray} 
where the following definitions are used:
\begin{equation}
\Delta I = \frac{8\pi}{15}\int_0^R \rho \frac{d}{dr}(\tilde{d}(r)r^5)dr,
\label{deltaI}
\end{equation} 
and
\begin{equation}
\tilde{d}(r_0)=-\frac{Gm}{R}\Bigl(\frac{r_0}{R}\Bigl)^2 q_t\frac{H(r_0)}{r_0 g(R)}.
\end{equation}
The total deformation factors $d_{ij}$ are given by a combination of static deformations ($\bar{d}_{ij}(r)$) and periodic deformations ($\tilde{d}(r)$).

The periodic components of the inertia tensor $I_{11}^p$ and $I_{22}^p$ are associated with $A$ and $B$ variations called radial tides and $I_{33}^p$ is associated with $C$ variations called zonal tides. The $I_{12}$ component corresponds to a periodic bulge shifted by $\pi/4$ in longitude from the principal axis of inertia (called librational tide)\citep{Murray99,Greenberg98}. 

\begin{figure}[!htbp]
 \centering
 \includegraphics[scale=0.24]{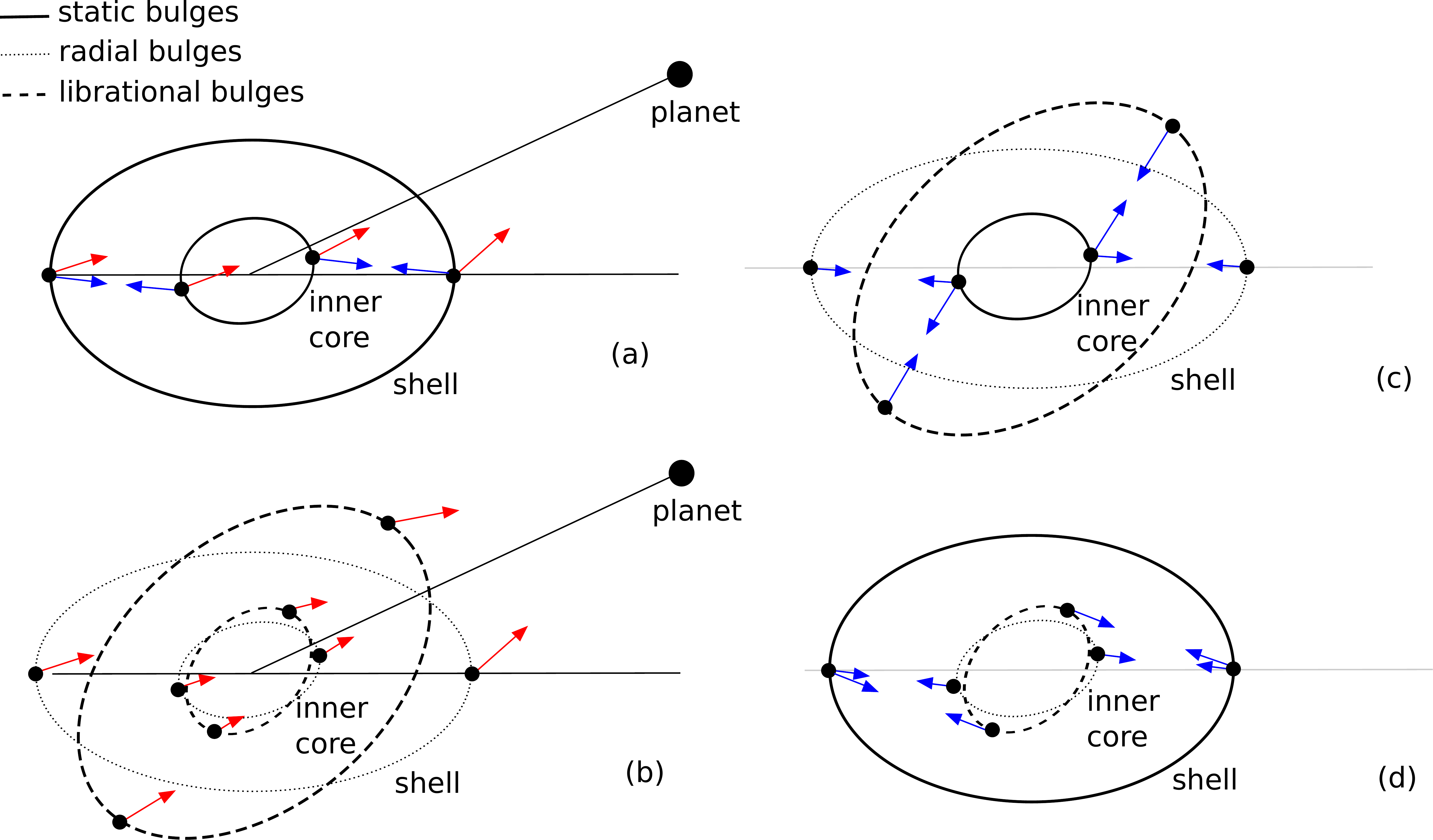}      
%% Note the ABSENCE of the extension .pdf , .eps or .ps  !
  \caption{Schematic position of the different forces acting on the bulges of the icy shell and the solid inner core when the satellite's mean anomaly $M$ is small and positive. The angle with the planet position and the radial and librational bulges amplitude have been exaggerated. Red arrows are the gravitational forces exerted by the planet and blue arrows are the internal gravitational forces between solid layers. (a) The static bulges are attracted by the planet and attracting each other due to misalignment. (b) The planet exerts gravitational forces on the radial and librational bulges. The resulting torques tend to oppose themselves in the radial and librational components of a layer. Internal gravitational forces are acting between periodic bulges and the static bulge of the core (c) and of the shell (d). The gravitational interaction between the periodic bulges is not represented.}
  \label{figure4}
\end{figure}

The external gravitational torque exerted on a layer $l$ is given by $\vec{\Gamma}_{ext}^l=\vec{r}\times M_p\vec{\nabla}V_{ext}^l$ where $V_{ext}^l$ is the external gravitational potential of the deformed satellite layer $l$ given by \citep{Jeffreys76}
\begin{multline}
V_{ext}^l(r_0,\theta,\lambda)=-\frac{4\pi G}{r_0}\int_{r^{l-1}}^{r^l}\rho(r)r^2dr\\-\sum_{j=0}^2\frac{4\pi G}{5r_0^3}\int_{r^{l-1}}^{r^l} \rho(r)\frac{d}{dr}(d_{2j}(r)r^5)drP_{2}^j(\cos \theta)\cos j\lambda\\-\frac{4\pi G}{5r_0^3}\int_{r^{l-1}}^{r^l} \rho(r)\frac{d}{dr}(e_{2j}(r)r^5)drP_2^j(\cos \theta)\sin j\lambda,
\end{multline}
where $r^l$ and $r^{l-1}$ are respectively the top and bottom radius of the layer $l$.
The z-component of this torque can be written at second order in eccentricity
\begin{multline}
\Gamma_{ext}^l=(K_l(1+3e\cos M) 
 +K_l^pe\cos M)(\nu-M-\gamma_l-\phi_{0,l})\\-\frac{2}{3}K_l^pe\sin M -K_l^pe^2\sin 2M\ .
\label{gammaext}
\end{multline}
Here $K_l$ describes the torque amplitude exerted by the planet on the static bulge of the layer $l$ as in the rigid case (see Figure \ref{figure4}(a)). $K_l^pe\cos M$ is the torque amplitude exerted on the radial tidal bulge of the layer $l$ with $K_l^p=\frac{9}{2} n^2 (\Delta I_l+\Delta I_l')$ and $\Delta I_l'$ is the contribution of the ocean pressure on the periodic figure based on equation (\ref{deltaI}). 

%\tar{As the static shape of the solid layers, the elastic deformations of the shell and the inner core induce a pressure on the ocean. This pressure is expressed by using the Navier-Stokes equation in the case of a Poincare flow (see e.g. \cite{Hinderer82}). It is then related to the different potentials applied to the ocean
%\begin{equation}
%\vec{\nabla}p=-\rho_o\vec{\nabla}V_{ext}-\rho_o\vec{\nabla}V_{int}+\rho_o\vec{\nabla}\Pi+\rho_o\vec{\nabla}\Psi,
%\end{equation}
%where $\rho_o$ is the ocean density, $p$ the pressure, $\Pi$ the inertial potential and $\Psi$ the centrifugal potential. Since the $z$-components of the centrifugal and inertial torques are negligible,  the pressure torque is expressed as a contribution to the gravitational torques and as variations to the moments of inertia. This approach is different from \cite{Goldreich10} who expressed the pressure torque by the mean of stress-strain relations and the stress tensor components.}

The last two terms of equation (\ref{gammaext}) are the torque exerted by the planet on the librational bulge. It can be notice that the development in second order of $K_l^pe\cos M(\nu-M-\phi_0)$ for a Keplerian orbit cancels out $K_l^pe^2\sin 2M$, and in this case the radial tides terms vanishes. The signs of the remaining librational and static terms at first order in $e$ are in phase and opposed, which means that the torques are counter-acting as illustrated by the gravitational forces on Fig. \ref{figure4}(b).

The total internal gravitational torque exerted by outer layers on the interior is given by $\vec{\Gamma}_{int}=-\int_i(\vec{r}\times\vec{\nabla}V_{int})\rho dV$ where $V_{int}$ is the potential exerted by outer layer on an inner point given by \citep{Jeffreys76}
\begin{multline}
V_{int}(r_0,\theta,\lambda)=-4\pi G\int_{r_i}^R\rho(r)rdr-\\\sum_{j=0}^2\frac{4\pi G}{5}r_0^2\int_{r_i}^R\rho(r)\frac{d}{dr}(d_{2j})dr P_{2}^j(\cos \theta)\cos j\lambda \\ -\frac{4\pi G}{5}r_0^2\int_{r_i}^R\rho(r)\frac{d}{dr}(e_{2j})drP_2^j(\cos \theta)\sin j\lambda.
\end{multline}
where $i$ refers to the inner core of mean radius $r_i$. The z-component of the internal gravitational torque exerted on the inner core is then
\begin{multline}
\Gamma_{int}=-(K_{int}+\frac{3}{2}K_{int}^{s/p}e\cos M+\frac{1}{4}K_{int}^{p/s}e\cos M)\sin 2(\phi_i-\phi_s) \\
- (2K_{int}^{s/p}-\frac{1}{3}K_{int}^{p/s})e\sin M \cos 2(\phi_i-\phi_s),
\label{gammaint}
\end{multline}
where $K_{int}^{s/p}$ and $K_{int}^{p/s}$ are defined such as
\begin{align}
K_{int}^{p/s}&=\frac{24\pi G}{5}\int_{r_i}^R \rho (r) \frac{d}{dr}(\tilde{d}(r))dr[(B_i-A_i)-(B_i'-A_i')],\\
K_{int}^{s/p}&=\frac{24\pi G}{5}\int_{r_i}^R \rho (r)\frac{d}{dr}(\bar{d}_{22}(r))dr(\Delta I_i-\Delta I_i'),
\end{align}
with $\bar{d}_{22}$ the deformation factor of the static bulge. In the internal gravitational torque, three main components are identified: the interaction between the static bulges of the shell and the inner core through $K_{int}$ like in the rigid case (see Figure \ref{figure4}(a)), the interaction between the periodic bulge of the shell and the static bulge of the core (subscript $p/s$, Figure \ref{figure4}(c)), and the interaction between the static bulge of the shell and the periodic bulge of the core (subscript $s/p$, Figure \ref{figure4}(d)). Here, the attraction between periodic bulges, which is of higher order in eccentricity, is neglected. Terms in $\cos M$ are due to radial tides while terms in $\sin M$ are due to librational tides. All the term but one have the same sign in the internal torque, which means that, for small and positive values of $M$, they are acting in the same direction and tend to align the shell and the inner core figures. The last term is counter-acting however, it corresponds to the attraction of the static bulge of the core by the librational bulge of the shell (Fig. \ref{figure4}(c)).

%===
% FIGURE TRANSFER FUNCTION
%==

The equations of librations are then linearized and written at second order in eccentricity:
\begin{equation}
\left\{
\begin{array}{ll}
C_s\ddot{\gamma_s}-n\frac{\Delta I_s}{2}(n+\dot{\gamma_s})e\sin M &= \Gamma_{ext}^s-\Gamma_{int},\\
C_i\ddot{\gamma_i}-n\frac{\Delta I_i}{2}(n+\dot{\gamma_i})e\sin M &= \Gamma_{ext}^i+\Gamma_{int},
\end{array}
\right.
\label{syslibdef}
\end{equation}
where the second term of the left-hand side is due to zonal tides on the layer $l$ with $\dot{C}_l=\dot{I}_{33}^{p,l}$. The internal and external gravitational torques at first order in eccentricity are dominated by the static and librational tides and the second order is dominated by the radial tides.

\subsection{Analytical resolution of the elastic case}
\label{sec:elastic}

The system (\ref{syslibdef}) is composed of two differential equations of second order in $\gamma_s$ and $\gamma_i$. To solve this system, we transform it into a first order equation system in $z$ by setting $z_1=\gamma_s$, $z_2=\gamma_i$, $z_3=\dot{\gamma}_s$, and $z_4=\dot{\gamma}_i$. The system is then decomposed into a static and a periodic part such as $\mathbf{\dot{z}}=\mathbf{Az}+\Delta\mathbf{A}(t)\mathbf{z}+\mathbf{b}(t)$ where $\textbf{z}$ is the vector of components $z_i$, $\mathbf{b}$ is the vector of forcing terms, $\mathbf{A}$ and $\Delta\mathbf{A}$ are the matrices of the static and periodic coefficients, respectively. The static coefficient matrix is diagonalized by defining a vector $\mathbf{y}$ such that $\mathbf{y}=\mathbf{P}^{-1}\mathbf{z}$ where $\mathbf{P}$ corresponds to the eigenvector matrix of $\mathbf{A}$. The system is then reduced to
\begin{equation}
\mathbf{\dot{y}}=\mathbf{\Lambda y}+\Delta \mathbf{\Lambda}(t)\mathbf{y}+\mathbf{f}(t)+\Delta \mathbf{f}(t),
\end{equation}
where $\mathbf{\Lambda}$ is the diagonal matrix, $\Delta\mathbf{\Lambda}$ is the transformation of $\Delta\mathbf{A}$ and $\mathbf{f}$ and $\Delta \mathbf{f}$ are the transformation of $\mathbf{b}$ decomposed into first order and higher order terms, respectively.

To solve this new system, the perturbation method used by \textit{e.g.} \cite{Robutel11} is followed and $\mathbf{y}$ is decomposed in decreasing amplitude terms such that $\mathbf{y}=\mathbf{y}_1+\mathbf{y}_2+...\ $ where subscripts $1$ and $2$ refer to first and second order solutions. We then have to solve
\begin{align}
\mathbf{\dot{y}}_1 &=\mathbf{\Lambda}\mathbf{y}_1+\mathbf{f}(t),\\
\mathbf{\dot{y}}_2 &=\mathbf{\Lambda}\mathbf{y}_2+\Delta\mathbf{\Lambda}(t)\mathbf{y}_1+\Delta\mathbf{f}(t), \label{2ndorder}\\ 
\vdots \nonumber 
\end{align}

The $\mathbf{y}_1$ system is first solved and the solutions are substituted in the $\mathbf{y}_2$ system to obtain the second order solutions. Then the solutions are transformed back to our libration angles also decomposed in $\gamma_{l,1}$ and $\gamma_{l,2}$. The solution $\mathbf{y}_3$ is not computed here since we are only interested in solutions of order 2 in eccentricity.

Since the deformation is also periodic with frequency $n$, the solutions $\gamma_{s,1}$ of the elastic case are given by 
\begin{eqnarray}
\gamma_{s,1,n}=\frac{1}{C_iC_s}\frac{H_n[(K_s-\Delta K_s)(K_i+2K_{int}-n^2 C_i)+2K_{int}(K_i-\Delta K_i)]}{(n^2 - \omega_1^2)(n^2 - \omega_2^2)},
 \label{gammaelastic}
\end{eqnarray}
where the third subscript corresponds to the frequency. At the orbital frequency, the amplitude is $H_n=2e$ and
\begin{align}
\Delta K_s &= \frac{e}{H_n}\Bigl(\frac{2}{3}K_s^p-2K_{int}^{s/p}+\frac{1}{3}K_{int}^{p/s}-n^2\frac{\Delta I_s}{2}\Bigl),\label{dKs}\\
\Delta K_i &= \frac{e}{H_n}\Bigl(\frac{2}{3}K_i^p+2K_{int}^{s/p}-\frac{1}{3}K_{int}^{p/s}-n^2\frac{\Delta I_i}{2}\Bigl). \label{dKi}
\end{align}
As suggested in the previous section, the torques variation $\Delta K_s$ and $\Delta K_i$ present in the solution at first order in eccentricity are only due to librational tides components.

The solution for an elastic solid body without ocean is given in Appendix.

\section{Librational response and application to interior models}~\label{sec:libration}
\subsection{Discussion of the rigid solution}
\label{sec:Discussion}
The libration amplitude obtained in the previous sections for the rigid and elastic cases are described and analyzed in this section. The libration responds to the perturbations according to three regimes depending on the values of the forcing frequencies with respect to the values of the proper frequencies: a high forcing frequency regime, a low forcing frequency regime, and a resonant regime where the forcing frequencies are close to the proper frequencies.

For all regimes, the solutions are simplified with the assumption that the icy shell thickness is small. $K_s/C_s$ is about 10 to 20 times larger than $\ K_i/C_i$ for the six models selected in Sect. \ref{sec:mis}. So the proper frequencies can be developed at first order in $\omega_i^2$ and $K_{int}/C_i$ such as

\begin{equation}
\omega_1^2  \sim \omega_s^2+\mu^2\ ,\label{propfreq1}
\end{equation}
\begin{equation}
\omega_2^2 \sim \omega_i^2\Bigl(\frac{\omega_s^2+2\frac{K_{int}}{C_s}}{\omega_s^2+\mu^2}\Bigl)+2\frac{\omega_s^2}{\omega_s^2+\mu^2}\frac{K_{int}}{C_i}\ .\label{propfreq2}
\end{equation}

Using these developments, it can be noticed that for low forcing frequencies $\omega_j^2 << \Bigl(\omega_i^2,\ \frac{K_{int}}{C_i}\Bigl)$, the libration amplitude of the rigid case is simply $\gamma_s \sim H_j$, i.e. the librational response of the body or the icy shell will follow the magnitude of the perturbation. In that case,  the librational response is dominated by the external gravitational torque and the internal structure has a negligible influence.

If the atmospheric torque excites the system then the amplitude of the librational response to low frequency forcing will be different from the magnitudes of the orbital perturbations. This difference would contain information on the atmospheric torque coupled to internal structure through the sine term
\begin{equation}
\gamma_{s,j}^s=H_j\Bigl(1+\frac{\Gamma_{A,j}}{H_jK_s}\cos(\Delta\alpha_j)(1-\frac{2K_{int}}{K_s+4K_{int}})\Bigl)\ .
\label{atmosolution}
\end{equation}
At low forcing frequency, the librational response due to the atmosphere is then inversely proportional to the external gravitational torque amplitude on the shell $K_s$ and dependent on the internal gravitational coupling through the constant $K_{int}$ which depends on the inner core composition.

At high forcing frequencies, the rigid solution (eq. (\ref{gamma})) can be simplified to
\begin{equation}
\gamma_{s,j}\sim-\frac{H_j\omega_s^2}{\omega_j^2-(\omega_s^2+\mu^2)}\ ,
%=\frac{H_j\Bigl[2(\omega_s^2+\omega_i^2)\frac{K_{int}}{C_i}+\omega_i^2\omega_s^2-\omega_j^2\omega_s^2\Bigl]}{(\omega_j^2-\omega_1^2)(\omega	_j^2-\omega_2^2)} 
%\label{F1}
\end{equation}
when $\omega_j^2$ is large in front of $\omega_i^2$ and $K_{int}/C_i$. Here, the dynamics is dominated by $\omega_s^2$ if $\omega_j^2>>(\omega_s^2+\mu^2)$. By writing $(B_s-A_s)$ as function of the equatorial flattening of the layers (see \cite{Vanhoolst09}), and by noticing that the flattening of the ocean is almost equal to the one of the shell from our interior models (the variations are only $2\%$), $\omega_s^2$ is written
\begin{equation}
\omega_s^2\sim 3n^2\beta_s\Bigl(\frac{\rho_o}{\rho_s}\frac{r_o}{5h}+1\Bigl),
\label{omegas}
\end{equation}
where the icy shell thickness $h$ and equatorial flattening $\beta_s$ have been introduced.
The free frequency and the libration amplitude are then dependent on the ratio between ocean and icy shell densities, and inversely proportional to the icy shell thickness.

Finally, if the forcing frequencies $\omega_j$ are close to the proper frequencies of the system, the librational response is dominated by the resonant behavior. The amplitude of the libration is then significantly increased due to the presence of a small divisor. If the libration amplitude becomes too large, the linearized equations used are not enough to describe the dynamics. For Titan, the proper frequencies listed in Table \ref{table3} compared to the forcing frequencies listed in Table \ref{table2} show that there is no close resonance between these frequencies.

\subsection{Discussion of the elastic solution}
\label{sec:discelas}
Here, we assess the effect of the elasticity on the libration solution which is dominant only for high forcing frequencies. The elasticity induces periodic variations of the torques amplitudes that are identified in the diurnal libration amplitude through two terms denoted by $\Delta K_s$ and $\Delta K_i$ (eq. \ref{gammaelastic})  corresponding respectively to the variation of the shell and the inner core inertia. The form of the elastic case diurnal solution at first order in eccentricity (eq. (\ref{gammaelastic})) is similar to the rigid case solution (eq. (\ref{gamma})) and it can also be simplified as
\begin{equation}
\gamma_{s,1,n}\sim-\frac{H_n\Delta{\omega_s}^2}{n^2-(\omega_s^2+\mu^2)},
\label{gammasnlim}
\end{equation}
with $n^2$ large in front of $(K_i-\Delta K_i)/C_i$ and $K_{int}/C_i$, and $\Delta{\omega_s}$ is defined as
$$\Delta{\omega_s}=\sqrt{\frac{K_s-\Delta K_s}{C_s}}.$$
Thus, the libration amplitude at orbital frequency will be largely reduced compared to the rigid case if $\Delta K_s$ is large enough in front of $K_s$, \textit{i.e.} if the shell librational tide contribution is large enough in front of the external gravitational torque exerted on the static bulge. To evaluate this contribution, the reduction rates of the torques amplitudes are defined as $F_s=\frac{\Delta K_s}{K_s}$ for the shell and $F_i=\frac{\Delta K_i}{K_i}$ for the inner core. Their numerical values are given in the next section.

The second order librational reponse $\gamma_{s,2,j}$ to the forcing frequency $\omega_j\ne n$ is identical to the rigid case and the behavior is described in the previous section.
Terms of frequencies $(n\pm\omega_j)$ also appear in the second order solution due to the modulation of the gravitational torque amplitude.
Since $\omega_j$ can be small compared to $n$, the low frequency forcing can contribute to the orbital frequency libration but these terms have small amplitudes below one meter. 
For the special case where $\omega_j$ is close to $n$, the terms of frequency $(n-\omega_j)$ contribute to the libration with their amplitude being inversely proportional to the proper frequencies (see eq. \ref{order2}).
The solution $\gamma_{s,2,(n-\omega_j)}$ can be written for the frequency $(n-\omega_j)<<1$, $K_i^p<<K_i$ and $H_j>>\gamma_{s,2,j},\gamma_{c,2,j}$
\begin{equation}
\gamma_{s,2,(n-\omega_j)}=-\frac{1}{2C_iC_s}\frac{H_je[K_i(K_s^p+3K_s)+2K_{int}(K_s^p+3K_s+K_c)]}{\omega_1^2\omega_2^2}.\label{order2}
\end{equation}In that case, since the values of proper frequencies $\omega_1$ and $\omega_2$ are small (see Table \ref{table3}), large values of magnitude $H_j$ can provide a non-negligible signature in the libration as detailed in the next section.

%===
% TABLE PROPEr FREQUENCIES
%==
\begin{table*}[!htbp]
\footnotesize
\caption{\label{table3} Proper frequencies for the different interior models selected in this study and presented in Table \ref{table1}. $\omega_f$ is defined in the Appendix.}
\begin{center}
\begin{tabular}{l c c c}
 \hline
 &\begin{bf} $\omega_1$ (rad day$^{-1}$) \end{bf}  &\begin{bf} $\omega_2$ (rad day$^{-1}$) \end{bf}  &  \begin{bf} $\omega_f$ (rad day$^{-1}$) \end{bf} \\
  \hline
  \begin{bf} F1 \end{bf} & $2.1476\ 10^{-2}$ & $6.5018\ 10^{-3}$ & - \cr
  \begin{bf} F2 \end{bf} & $2.2381\ 10^{-2}$ & $5.7250\ 10^{-3}$ & -\cr
  \begin{bf} F3 \end{bf} & - & - & $7.3722\ 10^{-3}$\cr
   \begin{bf} CA10 \end{bf} & $2.2338\ 10^{-2}$ & $5.5170\ 10^{-3}$ & -\cr
    \begin{bf} MC11 \end{bf} & $2.2146\ 10^{-2}$ & $5.4227\ 10^{-3}$ & -\cr
     \begin{bf} FE10 \end{bf} & $2.1544\ 10^{-2}$ & $5.6297\ 10^{-3}$ & -\cr
        \hline
\end{tabular}
\end{center}
\end{table*}

\subsection{Application}
\label{sec:Application}

Now, the librational solutions are applied to the different interior models introduced in Sect.~\ref{sec:mis}. The librations amplitudes projected onto Titan's equator are given in Table \ref{table4}. As discussed previously the librations can be classified into three regimes depending on the value of the forcing frequency with respect to the values of the proper frequencies. 

For the low forcing frequencies, which are lower than the proper frequencies, the libration amplitude is almost equal to the magnitude of the orbital perturbation for any interior models. As seen in Table \ref{table4}, the resulting libration amplitudes are large, around $552$ meters and $470$ meters for the Saturnian semi-annual and annual frequencies. The small differences in the amplitudes at Saturnian semi-annual and annual frequencies are due to a second order effect in the proper frequencies (eqs. (\ref{propfreq1}) and (\ref{propfreq2})) and are related to the different inertia of the models, with a larger value of libration for the solid model induced by the small value of proper frequency $\omega_f$.

%===
% TABLE LIBRATIONAL AMPLITUDE OF THE ICY SHELL
%==
\begin{table}[!htbp]
\footnotesize
\caption{\label{table4}  Analytical icy shell libration amplitudes of an elastic Titan (as equatorial deviation in meters) for different internal structure models with a $100$ kilometers icy shell thickness and different forcing frequencies. The second part of the table is the resulting librations under gravitational and atmospheric torques, with the given amplitude TO05 of \cite{Tokano05} and a recent value CH12 derived from \cite{Charnay12} at Saturnian semi-annual frequency. The last part of the table is the libration amplitudes for models where solid layers are rigid. The libration amplitudes are truncated at $10$ meters.}
\begin{center}
\begin{tabular}{l l c c c r}
 \hline
 & & \multicolumn{3}{c}{\begin{bf}Amplitudes (m) \end{bf}}\cr
 \hline
  \begin{bf}Freq.  \end{bf}&\begin{bf} Period \end{bf}  &\begin{bf} F1 \end{bf}  &  \begin{bf} F2 \end{bf} &  \begin{bf} F3 \end{bf} & \begin{bf} Identification \end{bf}\\
    \begin{bf}(rad/days) \end{bf}&\begin{bf}(days)\end{bf}  &  &   &   & \\
  \hline
0.394018&      15.946441&  -62.061&  -86.257&  -50.661& $L_6-\varpi_{6}$\cr
       0.001169&    5376.6317& 552.772&  552.309&  560.119& $2L_{s}$\cr
      0.000584&   10750.4115&  470.287&  470.183&  471.838&  $L_{s}$\cr
      0.001753&    3584.9304&   72.120&   72.000&   74.306&  $3L_s$\cr
             0.000063&    99027.4111&  47.477 & 50.171 & 31.500 & $2\varpi_8-2\Omega_6-\varpi_6$\cr
                    0.001121&	5606.2511&	26.687 & 28.182 & 17.978 & $2L_s-\Omega_6-\varpi_6$\cr
      0.009810&     640.4878&   11.868&   16.118&  -24.275&    -\cr
        \hline\cr
0.001169&    5376.6317&  517.761&  515.967&  523.095&  Tokano 05\cr
        0.001169&    5376.6317& 548.164&  548.107&  556.305&   Charnay 12\cr
        \hline\cr
0.394018&      15.9464& -319.036& -384.527&  -52.032& Rigid case\cr
\\ 
         \hline
 & & \multicolumn{3}{c}{\begin{bf}Amplitudes (m) \end{bf}}\cr
 \hline
  \begin{bf}Freq.  \end{bf}&\begin{bf} Period \end{bf} & \begin{bf} CA10 \end{bf} &  \begin{bf} MC11 \end{bf} &  \begin{bf} FE10 \end{bf} & \begin{bf} Identification \end{bf}\\
    \begin{bf}(rad/days) \end{bf}&\begin{bf}(days)\end{bf}  &  &   &   & \\
  \hline
0.394018&      15.946441&   -72.751&  -76.679&  -75.420 &$L_6-\varpi_{6}$\cr
       0.001169&    5376.6317&  552.265&  552.284&  552.478  &$2L_{s}$\cr
      0.000584&   10750.4115&    470.171&  470.174&  470.217 &$L_{s}$\cr
      0.001753&    3584.9304&     71.994&   72.002&   72.054& $3L_s$\cr
      0.000063&    99027.4111& 51.963  & 51.996 & 50.714 & $2\varpi_8-2\Omega_6-\varpi_6$ \cr
             0.001121&	5606.2511& 29.185 & 29.205 & 28.492 & $2L_s-\Omega_6-\varpi_6$ \cr
      0.009810&     640.4878&      16.956&   17.351&   16.721& -\cr
        \hline\cr
0.001169&    5376.6317&    515.887&  515.995&  516.814& Tokano 05\cr
        0.001169&    5376.6317&   548.084&  548.055&  548.028&  Charnay 12\cr
\hline\cr
0.394018&      15.9464&  -391.131& -387.614& -358.397&Rigid case\cr
\hline
\end{tabular}
\end{center}
\end{table}

In the case of rigid layers, the librational response at orbital frequency are expected to be strongly dependent on the satellite inertia \citep{Vanhoolst09}. Differences between oceanic and solid models would then appear clearly through the variation of the amplitude by at least a factor six, as for example the values of 391 meters for CA10 and 52 meters for F3. In the rigid case, equation (\ref{omegas}) shows that the amplitude would mainly depend on the ocean and icy shell densities ratio and on the icy shell thickness.

As described in \ref{sec:Discussion}, the total torque acting on the solid layers is modified by reduction rates $F_s$ and $F_i$ when elasticity is taken into account. To evaluate these rates, the radial function $H(r)$ is computed by the numerical integration of the gravito-elastic equations set of \cite{Alterman59} with the SatStress numerical code developed by \cite{Wahr09}. For the different interior models selected, the corresponding surface Love numbers $h_2$ are contained between $1.3$ and $1.5$ for oceanic models, and about $6\times10^{-2}$ for the solid model. In an elastic and solid Titan, only a small deformation is necessary for the shear stresses to balance the tidal force and thus a smaller value of $h_2$. There are no shear stresses in an inviscid fluid layer allowing the Love number value to increase \citep{Rappaport08}. For the oceanic models, $H(r)$ values obtained in the shell are at least forty times larger than at the surface of the core. It means that the periodic radial deformation of the core is negligible compared to the shell as well as its radial and librational bulges. Thus the internal torque amplitude $K_{int}^{s/p}$ can be ignored in the forcing terms. 

For the CA10 model, we find $F_s\sim 81\%$ and $F_i\sim-63\%$, which means that the torque exerted on the shell is reduced by a factor $4/5$ while the torque exerted on the inner core is increased by a factor $2/3$. This opposition can be explained by the fact that the librational bulge of the shell is attracted by the inner core static bulge and the planet, and these forces are opposed to the motion of the shell static bulge. On the contrary, the librational bulge of the shell is attracting the static figure of the inner core and tends to increase its rotation angle. The libration amplitude of the CA10 model goes from $319.131$ m for the rigid case to $72.751$ m for the elastic case, which is a reduction of about $81\%$, \textit{i.e.} the reduction rates of the icy shell torque as expected from equation (\ref{gammasnlim}).

For the model without ocean (F3), the introduction of elasticity decreases libration amplitude from $52.032$ meters to $50.661$ meters, \textit{i.e.} a reduction of about $2\%$. This can be explained by the small deformation of an entire solid body, \textit{i.e.} the low value of $h_2$ at the surface. As a consequence, the introduction of the elasticity in the interior models reduces the diurnal libration amplitude of the oceanic models to the same order than the entire solid model libration. The distinction between oceanic and oceanless models is then much more difficult at orbital frequency.

The gravitational torque exerted on the deformed surfaces provides contributions in libration at frequencies $(n\pm\omega_j)$ that do not exist in a rigid system. Especially, as stated in Sect. \ref{sec:discelas}, the terms of frequency $(n-\omega_j$) with $\omega_j$ very close to $n$ provide a non-negligible amplitude. The computation of eq. (\ref{order2}) with forcing frequencies $\omega_j=0.394081$ rad days$^{-1}$ (due to interaction with Iapetus with the argument $L_6-2\varpi_8+2\Omega_6$) and $\omega_j=0.392897$ rad days$^{-1}$  (due to the motion of Saturn around the Sun with argument $L_6+\Omega_6-2L_s$) gives amplitudes of $51.9$ and $29.2$ meters respectively for the CA10 model. The amplitudes of these terms are $31.5$ and $18.0$ meters for the solid model F3, a difference due to the smaller radial deformations of the model. These librational terms contain the information on the presence of the subsurface ocean.  

%===
% FIGURE LIBRATION VS TIME
%==
 \begin{figure}[!htbp]
 \centering
\resizebox{\hsize}{!}{
 \includegraphics[width=0.6\textwidth,clip]{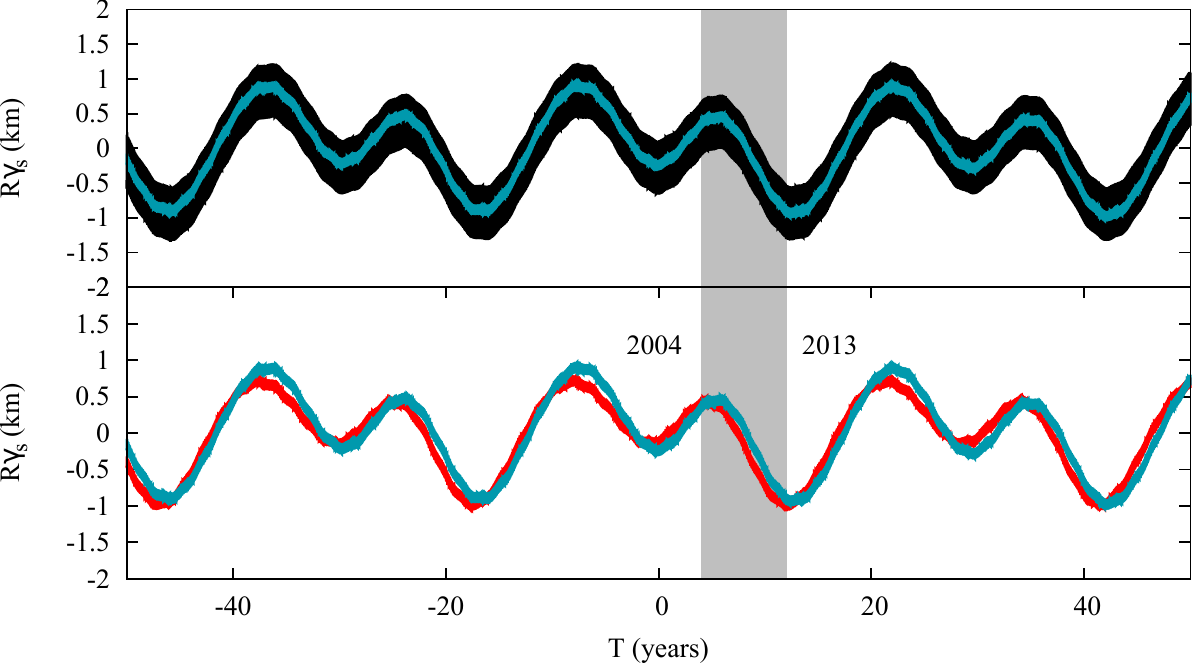}  }    
%% Note the ABSENCE of the extension .pdf , .eps or .ps  !
  \caption{Principal axis deviation of Titan's icy shell over 100 years for the CA10 internal structure model. Dark plot is the libration angle of the CA10 model with rigid solid layers, blue plot is the angle for the model with elastic solid layers and red plot contains elastic solid layers and the atmospheric excitation from \cite{Tokano05} of the Saturnian semi-annual frequency component. Initial date is J2000. The light gray band represents the actual Cassini mission lifetime.} 
  \label{figure2}
\end{figure}

The total icy shell librational response for the interior model CA10 is shown in Fig. \ref{figure2}. 
As described previously, the libration angle variation is dominated by the long period terms. The contribution of all amplitudes reaches $1$ kilometer at maximum, which could be detectable by radar  observations taken over $9$ years, represented in the gray box.
The short period librations are also visible in the thickness of the curve. The influence of elasticity is clearly visible in the diurnal libration amplitude (blue line) which is strongly
diminished with respect to the diurnal libration amplitude for
the rigid case (dark line). 

The presence of the atmospheric torque leads to additional terms in the Saturnian semi-annual frequency solution (eq. (\ref{atmosolution})) that are proportional to $\Gamma_{A,j}/K_s$, \textit{i.e.} the ratio of the torque magnitude to the constant $K_s$ describing the gravitational torque amplitude on the icy shell. The maximum of ${\Gamma_{A,j}}$ is obtained at the Saturnian semi-annual period with an amplitude of $1.6\ 10^{17}$ kg m$^2$ s$^{-2}$ for the atmospheric torque TO05. From the orbital analysis, the magnitude of the orbital perturbation $H_j$ at this frequency is equal to $2.1\ 10^{-4}$ rad and the coupling $K_s$ is $1.7\ 10^{21}$ kg m$^2$ s$^{-2}$ for the CA10 model. Then the ratio between atmospheric and gravitational coupling is $\frac{\Gamma_A}{H_jK_s} \sim 0.4$.
The atmospheric contribution to the solution is  then pondered by a factor $0.4\cos(\Delta\alpha)$ when the departure of the phases is taken into account, and $\Delta\alpha=-1.94$ rad at J2000 with the atmospheric torque TO05 described by \cite{Karatekin08}. The contribution is then of about ten percent, and even one order smaller with the atmospheric torque CH12 ($\Gamma_A=2.0\ 10^{16}$ kg m$^2$ s$^{-2}$ and $\Delta\alpha=-1.75$ rad). Other frequencies present in the atmospheric torque, like the Saturnian annual one, possess magnitudes lower than $5\ 10^{15}$ kg m$^2$ s$^{-2}$, leading to variations of the libration angle below a meter.

The red curve of Fig. \ref{figure2} presents the librational response of an elastic Titan with the atmospheric torque TO05. 
The main influence of the TO05 atmospheric torque is to introduce a shift in the librational motion and to thus slightly reduce the amplitude of the libration from $560$ meters to about $516$ meters, i.e. $8\%$. 
The shift is due to the phase difference with the orbital perturbation $\Delta\alpha$ of $111$ degrees. 
Using the CH12 atmospheric torque which has a ten times smaller Saturnian semi-annual frequency torque amplitude, the libration is only reduced to $548$ meters (with a phase shift $\Delta\alpha$ of $100$ degrees). In that case, the librational motion is very close to the libration without atmospheric torque and the librational motion of Titan results mainly from the orbital perturbations, with a variation of $5.5$ \% due to the atmosphere.

%===
% TABLE LIBRATIONAL AMPLITUDE OF THE ICY SHELL
%==
\begin{table}[!htbp]
\footnotesize
\caption{\label{table50km} Amplitudes of icy shell libration of elastic Titan's CA10 model (as equatorial deviation in meters) with a $50$, $75$ and $100$ km thick icy shell. The ocean density has been adjusted to conserve the mass and inertia of Titan. The amplitude of the librations are truncated at $10$ meters. The second part of the table includes the atmospheric coupling and the third part is the solution for a rigid case. }
\begin{center}
\begin{tabular}{l r c c c r}
 \hline
  \begin{bf}Freq.  \end{bf}&\begin{bf} Period \end{bf}  &\multicolumn{3}{c}{\begin{bf}Amplitudes (m) \end{bf}} &\begin{bf} Ident. \end{bf} \\
   \begin{bf}(rad/days) \end{bf}&\begin{bf} (days)\end{bf}  & $h=100$ km   &$h=75$ km   &  $h=50$ km  &   \\
  \hline
0.394018&      15.9464& -73.542& -73.770 & -61.607 & $L_6-\varpi_{6}$ \cr
       0.001169&    5376.6317&  552.265 & 551.821 & 551.178 & $2L_{s}$ \cr
       0.000584&   10750.4115& 470.171& 470.079 & 469.950 & $L_{s}$ \cr
       0.001753&    3584.9304&  71.994&   71.857&  71.648 &  $3L_s$ \cr
       0.000063&    99027.4111&  51.963 & 52.818 &  54.152 & $2\varpi_8-2\Omega_6-\varpi_6$ \cr
       0.001121&	5606.2511&	29.185 & 29.643 &	30.367 & $2L_s-\Omega_6-\varpi_6$ \cr
       0.009810&     640.4878&  16.956 & 15.762 &  14.123 & - \cr
        \hline \\
        
   0.001169&    5376.6317& 515.887 & 515.487 & 515.024 & Tokano 05 \cr
   0.001169&    5376.6317& 548.084 & 547.635 & 547.177 & Charnay 12\cr
   \hline\\
   0.394018&      15.9464& -391.131 &-507.087 & -736.871 & Rigid case \cr
   \hline
\end{tabular}
\end{center}
\end{table}

The icy shell thickness has been taken here as the maximum value predicted by observation, \textit{i.e.} $100$ kilometers. An icy shell thickness of $50$ kilometers would increase the diurnal libration amplitude of a rigid satellite by about a factor $2$ as shown in Table \ref{table50km} for a modified CA10 model. The radial function $H(r)$ is also larger of about $6\%$, and the reduction rate $F_s$ is equal to $91\%$. With a libration amplitude of $736$ m for rigid layers, the elastic libration with a $50$ km icy shell is then of about $62$ m, a value close to the rigid case. The icy shell thickness has almost no influence on the diurnal libration amplitude for a satellite with elastic solid layers in agreement with \cite{VanHoolst13}.

\subsection{Numerical integration}
\label{sec:num}

In parallel to the analytical theory, the non-linear librational equations are integrated numerically by taking into account the external and internal gravitational couplings
\begin{equation}
\left\{
\begin{array}{ll}
\frac{d^2 C_s\phi_s}{dt^2} &= \Gamma_{ext}^s-\Gamma_{int},\\
\frac{d^2 C_i \phi_i}{dt^2} &= \Gamma_{ext}^i+\Gamma_{int}.
\end{array}
\right.
\label{syslibdefnum}
\end{equation}
Each rotational angle $\phi_l$ is related to the librational angle through the relation $\phi_l = M + \gamma_l$. The orbital perturbation spectrum has been described in Section \ref{sec:orbit} and we explore numerically only the interior model CA10 described in Section \ref{sec:mis} with elastic solid layers. Numerical solutions of the libration angles of the shell $\gamma_s$ and of the inner core $\gamma_i$ are obtained. 

The frequency analysis method is applied on these numerical solutions in order to extract the amplitude and frequency of each term. The librational results are listed in Table~\ref{tab:numeric} for the icy shell. 

The comparison of the numerical amplitudes with those of analytical solution (Table \ref{table4}) shows very good agreement. The two terms (lines 5 and 6) due to interaction with Iapetus and the Sun are the terms of second order in eccentricity given by eq. (\ref{order2}). The small amplitude differences of the other terms is ascribed to the second order approximation in the analytical theory.

\begin{table}[!h]
\footnotesize
\caption{Frequency analysis of the numerical solution for the CA10 interior model with elastic solid layers.}
\begin{center}
\begin{tabular}{ l c c c }
 \hline
\begin{bf}Freq. \end{bf}&\begin{bf} Period \end{bf}  &\begin{bf} Amplitude \end{bf}  &  \begin{bf} Phase \end{bf} \\
\begin{bf}(rad/days) \end{bf}&\begin{bf} (days)\end{bf}  &\begin{bf} (m) \end{bf}  &  \begin{bf} (degree) \end{bf}  \\
  \hline
0.001169 &   5374.8365 &    552.182 &    -66.023 \cr 
0.000584 &  10758.8791 &    470.257 &    138.483 \cr 
0.394018 &     15.9464 &     74.036 &    -16.628 \cr 
0.001753 &   3584.2472 &     72.003 &    250.141 \cr 
 0.000063 &  99590.2556 &     51.704 &     128.468 \cr
 0.001120 &   5605.5072 &     29.068 &     -27.509 \cr
0.009810 &    640.4878 &     16.958 &    -77.290 \cr 
\hline
\end{tabular}
\end{center}
\label{tab:numeric}
\end{table}%

\section{Conclusion}

In this paper, the librational response in longitude of Titan has been investigated by taking into account the presence of a global ocean, the atmospheric torque, the orbital perturbations and the elasticity of solid layers. A linearized model has been developped and the amplitudes of libration have been computed and decomposed into responses at different frequencies. The librational responses can be divided into three regimes depending on the forcing frequencies: the low forcing frequencies at Saturnian annual and semi-annual periods, the high forcing frequencies at Titan's orbital and semi-orbital periods, and the forcing frequency ranges close to the proper frequencies of the system.

For the forcing at low frequencies, the libration response of the satellite is dominated by the gravitational torque and the internal structure has a negligible effect. In the high frequency forcing case, the librational response depends on the inertia of the body. The rigid case presents a large librational response at orbital frequency that is strongly reduced by the introduction of elasticity as stated by \cite{Goldreich10}. 
They introduced an elastic torque that corresponds to the pressure exerted by the ocean on the deformed icy shell as explained by 
\cite{Goldreich10} in Appendix (A.6). In our approach, the expression of the pressure torque is obtained according to the procedure detailed in \cite{Vanhoolst09}. The terms that result from the pressure torque are represented with a prime as shown in section 4.1 for the rigid case and section 4.2 for the purely non-rigid case. The pressure torque for the non-rigid case results from the deformation of the shell-ocean interface and it is characterized by the term $9/2 n^2 \Delta I'_l$.  
Thus physically, the pressure torque $9/2 n^2 \Delta I'_l$ and the pressure torque of \cite{Goldreich10} have the same origin. Then, the literal comparison of the two expressions is difficult because the two approaches are different. \cite{Goldreich10} assumed that the shell is a massless membrane overlaying a purely fluid body. In our case, Titan is divided in three layers, a shell of thickness 100 km, an ocean of about 200 km thick, and a solid core and the Love numbers formalism is used to compute the elastic deformation.

The internal structure of Titan is still under debate \citep{Castillo10,Castillo12, McKinnon11, Fortes12}. The six different interior models selected here (see Table \ref{table1}) are representative of the main categories of Titan's interior models. The librations computed for each model show that the libration amplitude for oceanic ($62-86$ m) or oceanless models ($51$ m) are of the same order at Titan's orbital frequency. We identify in the librations' Fourier decomposition two new frequencies resulting from the time-dependent inertia tensor in the external gravitational torque. The amplitudes of these librations are different for interior models with or without ocean. A reduction from $52$ to $30$ meters is obtained for the first term, and from $29$ to $18$ meters for the second one. These two long-period terms are interesting because they contain the information on the satellite internal structure.

Here, icy shell thicknesses of $100$, $75$ and $50$ kilometers have been investigated with a resulting libration amplitude at orbital frequency of the same order ($73.5$, $73.7$ and $61.6$ m, respectively). As shown by \cite{VanHoolst13}, the sensitivity of librations is less dependent on the ice shell thickness when elasticity of the solid layers is taken into account.

The atmospheric torque CH12 obtained with a recent General Circulation Model from \cite{LB12} presents a Saturnian semi-annual component ten times smaller than in the torque TO05 from \cite{Tokano05}. At this frequency, the orbital perturbation induces a large librational response of about $560$ meters, and the atmospheric torque TO05 contributes for about $10\%$ of the libration amplitude. The contribution is reduced to a few percent for the atmospheric torque CH12.

The librational response at Saturnian semi-annual frequency contain a small signature of the atmospheric circulation and the inner core inertia as shown in eq.(\ref{atmosolution}). In addition, we found that for a Titan containing an internal ocean the elasticity strongly reduces the amplitude of high frequency librations and the resulting amplitude is at the same order as the amplitude for solid Titan. Finally we identified two new librations (with arguments $2\varpi_8-2\Omega_6-\varpi_6$ and $2L_s-\Omega_6-\varpi_6$ ) at long period induced by the elasticity that contain a signature of the internal structure of a few ten meters. It is challenging to distinguish these small amplitudes librations with periods of about 30 and 50 years.

\appendix
\section{Solid case}
\label{sec:ann}

In the solid case, the system (\ref{eqlib}) reduces to one equation
\begin{equation}
C\ddot{\gamma}+3n^2(B-A)\gamma=3n^2(B-A)(\nu-M-\phi_o),
\end{equation}
with $A$, $B$ and $C$ the principal moments of inertia of Titan, $\gamma$ is the libration angle, $M$ the mean anomaly,  $\nu$ is the true longitude, $\phi$ is the rotation angle of the satellite's longest axis measured from the line of the ascending node and $\phi_{o}$ its initial value, and $n$ is the mean motion.

In the solid case, the proper frequency and libration amplitude are given by
\begin{equation}
\omega_f = n\sqrt{\frac{3(B-A)}{C}}, \label{omegaf}\\
\end{equation}
\begin{equation}
\gamma = \frac{H_j\omega_f^2}{\omega_f^2-\omega_j^2}\ .
\end{equation}
where $H_j$ is the magnitude of the perturbation of frequency $\omega_j$.

When $\omega_j$ is small in front of $\omega_f$, the librational response tends towards $H_j$. On the other hand, when the forcing frequency is large in front of the proper frequency, the amplitude of libration is reduced with respect to $H_j$.

The atmospheric torque induces a librational response on the following form 
\begin{equation}
\gamma_f=\gamma_j^s\sin(\omega_j t+\alpha_j) +\gamma_j^c\cos(\omega_j t+\alpha_j)\ ,
\end{equation}
where
\begin{equation}
\gamma_{j}^s = \frac{H_j\omega_f^2+\frac{\Gamma_{A,j}}{C}\cos(\Delta\alpha_j)}{\omega_f^2-\omega_j^2}\ ,\\
\end{equation}
\begin{equation}
\gamma_c = \frac{\frac{\Gamma_{A,j}}{C}\sin(\Delta\alpha_j)}{\omega_f^2-\omega_j^2}\ ,
\end{equation}
with $\alpha_j$ the phase of the orbital perturbation and $\Delta\alpha_j$ the difference of phase with the atmospheric torque of magnitude $\Gamma_{A,j}$.

By including the elasticity, the equation governing the librational dynamics is changed at second order in eccentricity to
\begin{multline}
C\ddot{\gamma}+(K(1+3e\cos M)+K^p e \cos M)\gamma-n\frac{\Delta I}{2}\dot{\gamma}e\sin M= -K^p e^2 \sin 2M\\+(K(1+3e\cos M)+K^pe\cos M)(\nu-M-\phi_o)-\Bigl(\frac{2}{3}K^p-n^2\frac{\Delta I}{2}\Bigl)e\sin M ,
\end{multline}
where $K^p=\frac{9}{2}n^2\Delta I$. By decomposing these equations on terms of decreasing amplitude as $\gamma=\gamma_1+\gamma_2+...$, we can solve the first equation given by
\begin{equation}
C\ddot{\gamma_1}+K\gamma_1=K(\nu-M-\phi_o)-\Bigl(\frac{2}{3}K^p-n^2\frac{\Delta I}{2}\Bigl)e\sin M.
\end{equation}
The solution $\gamma_1$ at orbital frequency $n$ is then written
\begin{equation}
\gamma_1=\frac{2e\Bigl(\omega_f^2-\frac{\Delta K}{C}\Bigl)}{n^2-\omega_f^2},
\end{equation}
with $\Delta K=\frac{1}{2e}\Bigl(\frac{2}{3}K^p-n^2\frac{\Delta I}{2}\Bigl)$.

The solution at second order in eccentricity $e$ is given at the frequency $(n-\omega_j)<<\omega_f$ and $H_j>>\gamma_1$ by
\begin{equation}
\gamma_{2,j}= -\frac{H_je}{2C}\frac{3K+K_p}{\omega_f^2}.
\end{equation}

The external torque exerted on the solid body figure is canceled when the dynamic Love number $h_2$ is equal to the fluid Love number $h_f$. The torque amplitude is given at the orbital frequency by $$\Gamma_{ext}=2eK-\frac{2}{3}eK^p=6en^2(B-A)-3en^2\Delta I.$$ Since $$\Delta I=\frac{8\pi}{15}\rho\tilde{d}(R)R,\ \text{with}\ \tilde{d}(R)=3\frac{M_p}{m}\Bigl(\frac{R}{a}\Bigl)^3h_2,$$ and $$(B-A)=\frac{16\pi}{5}\rho \bar{d}_{22}R^5,\ \text{with}\ \bar{d}_{22}=\frac{1}{4}\frac{M_p}{m}\Bigl(\frac{R}{a}\Bigl)^3h_f,$$ we find $\Gamma_{ext}=0$ when $h_2=h_f$. In that case the torque exerted by Saturn on the satellite is null since the solid layers are responding to tidal forces as an incompressible fluid. This behavior is also explained by \cite{VanHoolst13} by the mean of the love number $k_2$.

\section*{Acknowledgements}
The authors wish to thank J. Laskar, J.L. Simon, S. Lebonnois, J. Castillo-Rogez, and P. Robutel for fruitful and valuable discussions on the different aspects of this work. The authors also thank the anonymous reviewers who helped to improve the quality of this work.

%\bibliographystyle{model2-names}
%\bibliography{mnemonic,biblio}

\end{document}